\documentclass[sigconf]{acmart}
\usepackage{booktabs}
\usepackage{amsmath}
\usepackage[linesnumbered, lined, ruled, vlined]{algorithm2e}
\usepackage{multirow}
\AtBeginDocument{%
  \providecommand\BibTeX{{%
    \normalfont B\kern-0.5em{\scshape i\kern-0.25em b}\kern-0.8em\TeX}}}

\setcopyright{acmcopyright}
\copyrightyear{2018}
\acmYear{2018}
\acmDOI{10.1145/1122445.1122456}

\acmConference[Woodstock '18]{Woodstock '18: ACM Symposium on Neural
  Gaze Detection}{June 03--05, 2018}{Woodstock, NY}
\acmBooktitle{Woodstock '18: ACM Symposium on Neural Gaze Detection,
  June 03--05, 2018, Woodstock, NY}
\acmPrice{15.00}
\acmISBN{978-1-4503-XXXX-X/18/06}

\begin{document}

\title{Vec2GC - A Simple Graph Based Method for Document Clustering}

\author{Rajesh N Rao}
\email{rajeshnagaraja.rao@in.bosch.com}
\affiliation{Robert Bosch Research and Technology Center, India
\city{Bangalore}
\country{India}}
\author{Manojit Chakraborty}
\email{manojit.chakraborty@in.bosch.com}
\affiliation{Robert Bosch Research and Technology Center, India
\city{Bangalore}
\country{India}}

\renewcommand{\shortauthors}{Rao and Chakraborty, et al.}

\begin{abstract}
  NLP pipelines with limited or no labeled data, rely on unsupervised methods 
  for text processing. Unsupervised approaches typically begin with clustering 
  of terms or documents. In this paper, we introduce a simple and novel clustering 
  algorithm, \textbf{Vec2GC} (Vector to Graph Communities), to cluster documents in a corpus. Our method uses community 
  detection algorithm on a weighted graph of documents, created using document 
  embedding representation. Vec2GC clustering algorithm is a density based 
  approach, that supports hierarchical clustering as well.
\end{abstract}




\maketitle

\section{Introduction}

Dealing with large collection of unlabeled domain specific documents is a 
challenge faced often in industrial NLP pipelines. Clustering algorithm 
provide a mechanism to analyze document collections when no information is 
available apriori. Document clusters provide hints on topics present in the 
corpus and a distribution of documents across different topics. Combined with 
visual representation based on dimensionality reduction we can get a good 
overview of data distribution.
\begin{figure}
	\centering
	\includegraphics[width=0.8\linewidth,height=6cm]{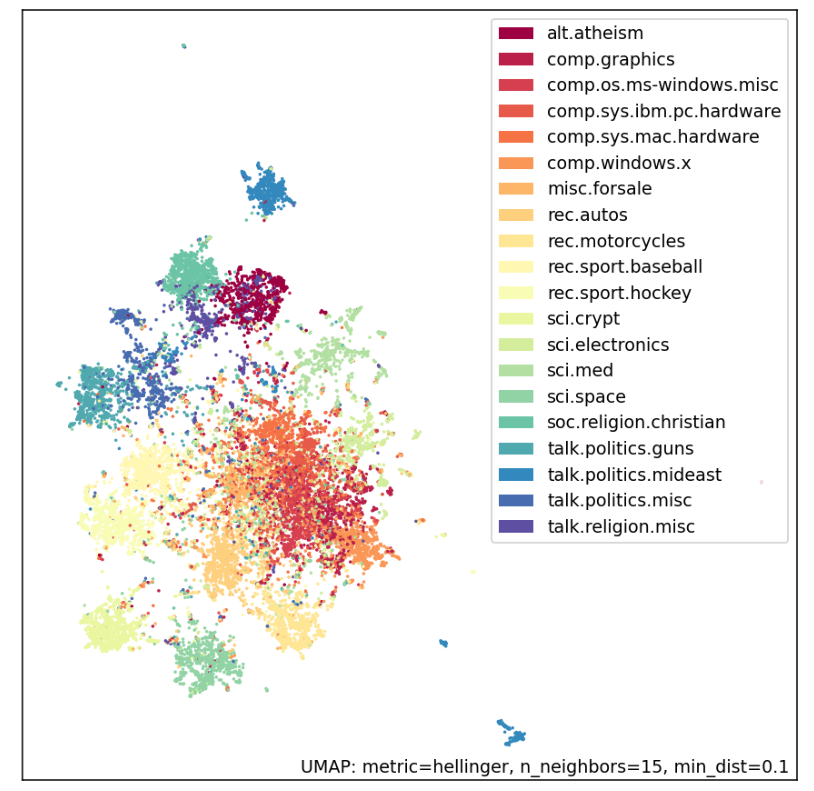}
	\caption{UMAP 2d plot of 20 Newsgroup Documents}
	\label{fig:20_ng_visualization}
\end{figure}
Dimensionality reduction techniques like PCA \cite{pearson}, t-SNE \cite{van} 
or UMAP \cite{umap} would map the document in embedding space to 2 dimensional 
space as shown in figure \ref{fig:20_ng_visualization}. Clustering based on 
document embeddings groups semantically similar documents together, to form 
topical distribution over the documents. Traditional clustering algorithms like 
k-Mean \cite{kmeans}, k-medoids \cite{saiyad}, DBSCAN \cite{dbscan} or HDBSCAN 
\cite{hdbscan} with distance metric derived from Cosine Similarity 
\cite{manning}, on document embeddings do identify clusters for topical 
representation, but we believe this can be improved. 

In prior works, Hossain and Angryk \cite{gdclust} represented text documents as 
hierarchical document-graphs to extract frequent subgraphs for generating 
sense-based document clusters. Wang et. al. \cite{wang-etal-2020-empirical} 
used vector representations of documents and run k-means clustering  on them to 
understand general representation power of various embedding generation models. 
Angelov \cite{angelov2020top2vec} proposed Top2Vec, which uses joint document 
and word embedding to find topic vectors representing dense regions in the 
embedding space identified using clustering method like HDBSCAN. Saiyad et. al. 
\cite{saiyad} presented a survey covering major significant works on semantic 
document clustering based on latent semantic indexing, graph representations, 
ontology and lexical chains. 

We propose the \textit{Vec2GC: Vector To Graph Communities}, a clustering 
algorithm that converts documents in the embedding space 
\cite{mikolov2013efficient} \cite{le2014distributed} to 
a weighted graph and generates clusters based on Graph Community Detection algorithm. Our contributions are as follows: 
\begin{itemize}
	\item We present a simple document clustering algorithm built from 
	combination of a non-linear weighted graph whose edge weights are derived 
	from cosine similarity of documents and standard graph community detection algorithm. 
	\item The algorithm is not restricted to any specific embedding 
	representation or to any specific Graph Clustering algorithm. Additionally, 
	Vec2GC provides a hierarchical density based clustering solution whose 
	granularity can be choosen based on application requirements. 
	\item We demonstrate that Vec2GC outperforms standard clustering 
	algorithms to provide better topical and semantic represenation of document clusters.
\end{itemize}

\section{Problem Statement}
Consider a document list $D = \{d_1, d_2, ... d_N\}$. We intend to identify 
clusters of documents denoted by cluster list $C = \{c_1, c_2, c_3, ... , 
c_M\}$. A document $d_i$ is assigned to either cluster $c_j$ or not assigned to 
any cluster. Documents that are not assigned to any clusters are considered as 
\textit{Noise}. 

Our goal is to assign as many documents as possible to non Noise clusters. 
Further more we desire the clusters created should make `semantic sense', i.e., 
the documents of a cluster should be topically similar. We demonstrate that 
Vec2GC out performs standard clustering algorithms in terms of topical 
similarity of cluster documents. 


\section{Vec2GC Algorithm}
Vec2GC converts the vector space embeddings of documents to a graph and executes a graph clustering on it. The two steps of Vec2GC algorithm are listed below:

\begin{itemize}
	\item Weighted graph construction from document embeddings.
	\item Hierarchical cluster generation from recursive use of Graph Community 
	Detection algorithm
\end{itemize}

\subsection{Graph Construction}
We consider each document as a node represented by $d_i$ and its embedding 
represented by $v_{d_i}$. To construct the graph, we measure the cosine 
similarity of the embeddings, equation (\ref{eqn:cosine_sim}). An edge is drawn 
between two nodes if their cosine similarity is greater than a specific 
threshold $\theta$, which is a tuneable parameter in our algorithm.

\begin{equation}
\label{eqn:cosine_sim}
cs (d_i, d_j) = \frac{v_{d_i} . v_{d_j}}{\parallel v_{d_i} \parallel \parallel 
v_{d_j} \parallel}
\end{equation}

The edge weight is determined by the cosine similarity value and is given by 
equation (\ref{eqn:edge_weight}).

\begin{equation}
\label{eqn:edge_weight}
E (d_i, d_j) =
\begin{cases} 
0 & cs(d_i, d_j) < \theta  \\
\frac{1}{1 - cs(d_i, d_j)} & cs(d_i, d_j) \ge \theta
\end{cases}
\end{equation}

Equation (\ref{eqn:edge_weight}) maps the cosine similarity to edge weight as 
shown below:

\begin{equation}
(\theta, 1) \rightarrow (\frac{1}{1 - \theta}, \infty)
\end{equation}
\\
As cosine similarity tends to 1, edge weight tends to $\infty$. 

Note in graph, higher edge weight corresponds to stronger connectivity. Also, 
the weights are non-linearly mapped from cosine similarity to edge weight. This 
increases separability between two node pairs that have similar cosine 
similarity. For example, a pair of nodes with $cs(a, b) = 0.9$ and another pair 
with $cs(x, y) = 0.95$ would have edge weights of $10$ and $20$ respectively. A 
stronger connection is created for cosine similarity closer to $1$. Thus higher 
weight is given to embeddings that are very similar to each other. 

Also, due to the threshold of $\theta$, only local neighboring nodes are 
considered. Only documents that have a cosine similarity greater than $\theta$ 
are connected.   

\subsection{Graph Community Detection}
The constructed graph consists of nodes representing documents and edges 
representing document pair with high cosine similarity. We apply a 
Graph Community detection algorithm to identify clusters in the graph. As 
mentioned in previous section, only document pairs with cosine similarity 
greater than $\theta$ are connected by an edge. The Graph Community detection 
algorithm would work to find nodes (i.e. documents) that are well connected 
with each other compared to other nodes. We apply the Graph Community Detection 
algorithm recursively to identify sub-communities of communities, till stopping 
criterion is satisfied.   

We apply a standard Graph Community Detection algorithm, \textbf{Parallel 
Louvian Method} \cite{blondel2008fast} to obtain a complete partition in 
determining the communities in the graph. We calculate the modularity index 
\cite{newman_modularity}, given by equation (\ref{eqn:modularity_index}), for 
each execution of the PLM algorithm. \\

\noindent
\textbf{Modularity Index : } Modularity Index provides a quantitative measure 
for the quality of clusters generated by the Graph Community Detection 
algorithm. Consider a complete partition that breaks a graph into ${N_c}$ 
communities. To see if the local link density of the subgraphs defined by this 
partition differs from the expected density in a randomly wired network, we 
define the partition's modularity index equation (\ref{eqn:modularity_index}).

\begin{equation}
\label{eqn:modularity_index}                                                 
Q = \frac{1}{2L} \sum_{a,b} \left[ W_{E_{ab}}  - \frac{k_a k_b}{2L}\right] 
\delta(c_a, c_b)               
\end{equation} 

\noindent
where $L$ is the number of edges in the graph, $W_{E_{ab}}$ represents the 
$(a,b)^{th}$ entry of the adjacency matrix $W$, $k_a$ is the degree of node 
$a$, $c_a$ is the label of the community to which node $a$ belongs to.

\vspace{2mm}
We execute the Graph Community Detection algorithm recursively. The pseudo code 
of the recursive algorithm in show in Algorithm \ref{alg:recur_gcd}
\begin{algorithm}
	\SetKwProg{Fn}{def}{}{}
	\BlankLine
	\Fn{GetCommunity($g,c\_node,tree,mod\_thresh,max\_size$)}{
		$mod\_index, c\_list = community\_detection\_algo(g)$\\
		\If{$mod\_index < \theta_{modularity}$}{
			tree.add\_node(curr\_node)\\
			return
		}
		\ForEach{$comm~~in~~c\_list$}{
			\If{$len(comm) > max\_size$}{
				s\_g = get\_community\_subgraph(comm)\\
				$n\_node = Node()$\\
				$tree.add\_node(n\_node)$\\
				$GetCommunity(s\_g, n\_node,tree, mod\_thresh, max\_size)$
			}
			\Else
			{
				$new\_node = Node()$\\
				$tree.add\_node(new\_node)$
			}
		}
	}
	\caption{Recursive Graph Community Detection\label{alg:recur_gcd}}
\end{algorithm}

\noindent
The $GetCommunity$ function internally calls the \\$community\_detection\_algo$ 
and $get\_community\_subgraph$ and itself recursively. The $GetCommunity$ 
function takes in a Graph, a \textit{modularity index threshold}, $m_d$, 
\textit{maximum community size} and an empty Tree data structure to be filled 
as output. The \\$community\_detection\_algo$ function gets the communities and 
modularity index for the given Graph. If the modularity index is below 
\textit{modularity index threshold}, communities are not well formed and are 
discarded. However if the modularity index is acceptable, the communities are 
created and added to the Tree data structure. If the newly formed communities 
are larger than the \textit{maximum community size}, a sub graph of the 
community members is created from $get\_community\_subgraph$ and passed to a 
recursive call to $GetCommunity$ function with updated Tree data structure. 

After the recursive execution of $GetCommunity$ ends, we get a Tree data 
structure that contains a hierarchical structure of communities in the Graph. 
We can select the hierarchy level to extract the communities. Documents that 
are not part of any communities are marked as \textit{Noise} and stored in a 
separate data structure.


\subsection{Noisy Nodes}
Note, not all nodes would be member of a community. There will be nodes that do 
not belong to any community. Nodes that are not connected or not closely 
connected fail to be a member of a community. We define such nodes as 
\textit{Noisy} nodes. We can modulate the number of Noisy nodes by changing the 
\textit{similarity threshold}, $\theta$ or the \textit{modularity index}, $m_d$.

\subsection{Datasets}
\paragraph{\textbf{20 newsgroups}}
The 20 Newsgroups data set comprises of approximately 20,000 newsgroup 
documents, evenly distributed across 20 different newsgroups, each 
corresponding to a different topic.
\footnote{\url{http://qwone.com/~jason/20Newsgroups/}}

\paragraph{\textbf{AG News}}
AG is a collection of more than 1 million news articles gathered from more than 
2000 news sources by ComeToMyHead 
\footnote{\url{http://groups.di.unipi.it/~gulli/AG_corpus_of_news_articles.html}},
which is an academic news search engine. The AG's news topic classification 
dataset is developed by Xiang Zhang \footnote{xiang.zhang@nyu.edu} from the 
above news articles collection of 127600 documents. It was first 
used as a text classification benchmark in the following paper \cite{zhang}

\paragraph{\textbf{BBC Articles}}
This dataset is a public dataset from the BBC, comprised of 2225 articles, each 
labeled under one of 5 categories: Business, Entertainment, Politics, Sport or 
Tech. \footnote{\url{https://www.kaggle.com/c/learn-ai-bbc/data}}

\paragraph{\textbf{Stackoverflow QA}}
This is a dataset of 16000 question and answers from the Stackoverflow website 
\footnote{\url{www.stackoverflow.com}}, labeled under 4 different categories of 
coding language - CSharp, JavaScript, Java, Python.
\footnote{\url{http://storage.googleapis.com/download.tensorflow.org/data/stack_overflow_16k.tar.gz}}

\paragraph{\textbf{DBpedia}}
DBpedia is a project aiming to extract structured content from the information 
created in Wikipedia. \footnote{\url{https://en.wikipedia.org/wiki/DBpedia}}
This dataset is extracted from the original DBpedia data that provides 
taxonomic, hierarchical categories or classes for 342,782 articles.
There are 3 levels of classes, with 9, 70 and 219 classes respectively. 
\footnote{\url{https://www.kaggle.com/danofer/dbpedia-classes/version/1}}

We use two different document embedding algorithms to generate document 
embeddings for all text datasets. The first algorithm that we use is \textbf{Doc2Vec}, 
which creates document embeddings using the distributed memory and distributed 
bag of words models from \cite{le2014distributed}. We also create document 
embeddings using \textbf{Sentence-BERT} \cite{reimers-2019-sentence-bert}. It computes 
dense vector representations for documents, such that similar document 
embeddings are close in vector space using pretrained language models on 
transformer networks like BERT \cite{devlin2019bert} / RoBERTa 
\cite{liu2019roberta}/ DistilBERT \cite{sanh2020distilbert} etc. in its 
framework. For our experiment, we use \textit{stsb-distilbert-base} 
\footnote{\url{https://huggingface.co/sentence-transformers/stsb-distilbert-base}} pretrained model, with Multi-head attention over 12 layers, max\textunderscore seq\textunderscore length = 128, word\textunderscore embedding\textunderscore dimension = 768, to generate document embeddings using Sentence-BERT.

To compare the effectiveness of our algorithm, we perform clustering on the 
document embeddings for each dataset using our proposed method \textbf{Vec2GC}, 
along with conventional document clustering methods \textbf{HDBSCAN} 
\cite{hdbscan} and \textbf{KMedoids} \cite{saiyad}. For KMedoids, we used an 
approach like KMeans++ as the medoid intialization method, which gives initial 
medoids which are more separated in vector space. For HDBSCAN, we used Excess 
of Mass algorithm as the cluster selection method to find the most persistent 
clusters. This gave use better result than Leaf method. HDBSCAN also creates a 
cluster labeled as -1, which contains noisy data points. We tuned the 
parameters of Vec2GC such that the number of data points in -1 cluster from 
HDSCAN matches approximately with the number of data points in the 
Non-Community Nodes community which we get as an output from Vec2GC, which also 
indicates noisy data points detected by Vec2GC, to maintain the experiments and 
comparisons unbiased.

\subsection{Results}

We perform cluster analysis with the results obtained from each of these 
methods. In our experiments we set the embedding vector size to $300$, 
$\theta=0.6$ and $m_d=0.3$. Cluster purity is a commonly used metric in 
cluster analysis to measure how good the clusters are. It measures the extent 
to which clusters contain a single class, or Homogeneity 
\cite{wang-etal-2020-empirical}. Here, we calculate purity for each cluster 
based on the document category. The number of data points from the most common 
class is counted for each cluster, for example, if the total number of data 
points in a cluster $C$ is 10, and the data points from the most common class 
in that cluster $C$ is 8, then cluster $C$ is said to have $(8/10)*100\% = 
80\%$  cluster purity. 

From the $N$ clusters obtained from a clustering method on a given dataset, we 
calculate the numbers of clusters that have 50\%, 70\% and 90\% purity, as 
$M_1,M_2,M_3$ respectively. Then we calculate the fractions $M1/N, M2/N, 
M3/N$. From the outputs of each clustering method (Vec2GC, HDBSCAN and 
KMedoids) on all five datasets, these three values are calculated individually 
and put into Table \ref{table:doc2vec_comparison} and Table 
\ref{table:sentence_bert_comparison}. Table \ref{table:doc2vec_comparison} 
results are from \textbf{Doc2Vec} document embeddings, where as Table 
\ref{table:sentence_bert_comparison} contains results from \textbf{Sentence-BERT} 
document embeddings. Best results are put in bold, second best results are 
underlined. 

\begin{table}[!htbp]
	\caption{Comparison using Doc2Vec Embeddings}
	\begin{tabular}{ |p{1.7cm}|p{1cm}|p{1.5cm}|p{1.5cm}|p{1.5cm}| } 
		\hline
		Dataset & Purity Value(k) & Fraction of clusters @ k\% purity (\textbf{KMedoids}) & Fraction of clusters @ k\% purity (\textbf{hdbscan}) & Fraction of clusters @ k\% purity (\textbf{Vec2GC}) \\
		\hline
		\multirow{3}{4em}{20Newsgroup} & 50\% & .53 & \underline{.76} & \textbf{.89}\\ 
		
		& 70\% & .38 & \underline{.56} & \textbf{.69}\\
		
		& 90\% & .07 & \underline{.20} & \textbf{.39}\\ 
		
		\hline
		\multirow{3}{4em}{AG News} & 50\% & \underline.98 & \underline{.98} & \textbf{.99}\\ 
		
		& 70\% & .74 & \underline{.90} & \textbf{.94}\\
		
		& 90\% & .20 & \underline{.63} & \textbf{.80}\\ 
		
		\hline
		\multirow{3}{4em}{BBC Articles} & 50\% & \textbf{1.0} & \underline{.99} & \underline{.99}\\ 
		
		& 70\% &  .86 & \underline{.93} & \textbf{.96}\\
		
		& 90\% & .50 & \underline{.70} & \textbf{.83}\\ 
		
		\hline
		\multirow{3}{4em}{DBPedia} & 50\% & .84 & \underline{.90} & \textbf{.93}\\ 
		
		& 70\% &  .52 & \textbf{.80} & \underline{.77}\\
		
		& 90\% &  .24 & \textbf{.54} & \underline{.53}\\ 
		
		\hline
		\multirow{3}{4em}{Stackoverflow} & 50\% & .30 & \underline{.63} & \textbf{.79} \\ 
		
		& 70\% & .14 & \underline{.35} & \textbf{.46}\\
		
		& 90\% & .02 & \underline{.15} & \textbf{.20} \\ 
		\hline
	\end{tabular}
	\label{table:doc2vec_comparison}
\end{table}

\begin{table}[!htbp]
	\caption{Comparison using Sentence-Transformer Embeddings (Using stsb-distilbert-base pretrained model)}
	\begin{tabular}{ |p{1.7cm}|p{1cm}|p{1.5cm}|p{1.5cm}|p{1.5cm}| } 
		\hline
		Dataset & Purity Value(k) & Fraction of clusters @ k\% purity (\textbf{KMedoids}) & Fraction of clusters @ k\% purity (\textbf{hdbscan}) & Fraction of clusters @ k\% purity (\textbf{Vec2GC}) \\
		\hline
		\multirow{3}{4em}{20Newsgroup} & 50\% & .46 & \underline{.64} & \textbf{.65}\\ 
		
		& 70\% & .27 & \textbf{.64} & \underline{.50}\\
		
		& 90\% & .09 & \textbf{.29} & \underline{.13}\\ 
		
		\hline
		\multirow{3}{4em}{AG News} & 50\% & .88 & \underline{.98} & \textbf{.99}\\ 
		
		& 70\% & \underline{.66} & \textbf{.90} & \textbf{.90}\\
		
		& 90\% & .18 & \textbf{.67} & \underline{.65}\\ 
		
		\hline
		\multirow{3}{4em}{BBC Articles} & 50\% &\textbf{ 1.0} & .94 & \underline{.98}\\ 
		
		& 70\% &  \textbf{.85} & .74 & \underline{.84}\\
		
		& 90\% & .30 & \underline{.47} & \textbf{.60}\\ 
		
		\hline
		\multirow{3}{4em}{DBPedia} & 50\% & .80 & \underline{.94} & \textbf{.99}\\ 
		
		& 70\% &  .54 & \underline{.88} & \textbf{.88}\\
		
		& 90\% &  .32 & \underline{.75} & \textbf{.77}\\ 
		
		\hline
		\multirow{3}{4em}{Stackoverflow} & 50\% & .13 & \underline{.28} & \textbf{.34} \\ 
		
		& 70\% & .05 & \underline{.10} & \textbf{.11}\\
		
		& 90\% & \underline{.01} & \underline{.01} & \textbf{.02} \\ 
		\hline
	\end{tabular}
	\label{table:sentence_bert_comparison}
\end{table}

As we can see from Table \ref{table:doc2vec_comparison} and 
\ref{table:sentence_bert_comparison}, for most of the datasets, Vec2GC clusters 
are the best with highest fraction of clusters with k\% purities. HDBSCAN comes 
second best for majority of the datasets, where as KMedoids gives the poorest 
clusters, in terms of cluster purity. This clearly shows that Vec2GC outperforms the baseline clustering methods for all datasets used and produces better semantic clusters. 

\paragraph{\textbf{Ablation Study}} We have studied the effect of introducing a non-linear function over cosine 
similarity to obtain edge-weights for the graph in our algorithm. Table 3 shows 
a comparison study of cluster purity between two methods: 1. Restricted Vec2GC, 
where we use cosine similarity values as edge-weights. 2. Vec2GC, where we use 
non-linear function derived from cosine similarity, as edge-weights described 
in Section 3. Results clearly show that when we use the non-linearity, 
cluster purity increases significantly and it is consistent over all the 
datasets used for the experiment. It empirically proves the notion of higher 
separability resulting in better clusters, which we described in section 3.1.

\begin{table}[!htbp]
	\caption{Effect of non-linearity in edge-weights}
	\begin{tabular}{ |p{1.7cm}|p{1cm}|p{1.5cm}|p{1.5cm}| } 
		\hline
		Dataset & Purity Value(k) & Fraction of clusters @ k\% purity (\textbf{Restricted Vec2GC}) & Fraction of clusters @ k\% purity (\textbf{Vec2GC}) \\
		\hline
		\multirow{3}{4em}{20Newsgroup} & 50\% & .55 & \textbf{.65}\\ 
		
		& 70\% & .37 & \textbf{.50}\\
		
		& 90\% & .07 & \textbf{.13}\\ 
		
		\hline
		\multirow{3}{4em}{AG News} & 50\% & .95 & \textbf{.99}\\ 
		
		& 70\% & .78 & \textbf{.90}\\
		
		& 90\% & .42 & \textbf{.65}\\ 
		
		\hline
		\multirow{3}{4em}{BBC Articles} & 50\% & .98 & \textbf{.98}\\ 
		
		& 70\% & .79 & \textbf{.84}\\
		
		& 90\%& .55 & \textbf{.60}\\ 
		
		\hline
		\multirow{3}{4em}{DBPedia} & 50\% & .98 & \textbf{.99}\\ 
		
		& 70\% & .86 & \textbf{.88}\\
		
		& 90\% & .74 & \textbf{.77}\\ 
		
		\hline
		\multirow{3}{4em}{Stackoverflow} & 50\% & \textbf{.35} & .34 \\ 
		
		& 70\% & .10 & \textbf{.11}\\
		
		& 90\%  & .01 & \textbf{.02} \\ 
		\hline
	\end{tabular}
	\label{table:sentence_bert_comparison}
\end{table}

\section{Conclusion and Future Works}

In this paper we discuss the Vec2GC algorithm that transforms a set of 
embeddings in a vector space to a weighted graph and recursively applies 
community detection algorithm to detect hierarchical clusters of documents. The 
Vec2GC algorithm leverages graph structure to capture local neighborhood of 
embeddings and also executes community detection algorithm recursively to 
create a hierarchical cluster of terms or documents. We experimented with 
different corpora and demonstrated that the Vec2CG clustering algorithm 
performs better than the standard clustering algorithms like k-mediods, DBSCAN 
or HDBSCAN, which are generally used in document clustering frameworks 
\cite{saiyad}. Our experiments demonstrate that for document embedding 
clustering, Vec2GC is a better clustering algorithm.

Currently we have shown the result of Vec2GC with respect to document 
clustering. However, this can be applied to terms as well. We will benchmark 
Vec2GC for clustering terms and compare it with existing clustering algorithms. 
Combining terms and documents in a single vector space provides an opportunity 
to create Topic Modeling clusters. Similar to \cite{angelov2020top2vec}, we 
intend to apply Vec2GC clusters to identify Topics in a given corpus.  


\bibliographystyle{ACM-Reference-Format}
\bibliography{reference}


\end{document}